\documentclass[11pt]{article}
\usepackage{graphicx}

\newcommand{\BABARPubYear}    {01}
\newcommand{\BABARProcNumber} {97}
\newcommand{\SLACPubNumber} {9059}

\input babarsym
\newcommand{\be}{\begin{equation}}
\newcommand{\ee}{\end{equation}}
\newcommand{\bd}{\begin{displaymath}}
\newcommand{\ed}{\end{displaymath}}
\newcommand{\bea}{\begin{eqnarray}}
\newcommand{\eea}{\end{eqnarray}}
\def\babar{\mbox{B\hspace{-0.4em} {\scriptsize A}\hspace{-0.4em} 
B\hspace{-0.4em} {\scriptsize A\hspace{-0.1em}R}}}

\def\FourS {\Y4S}

\def\Y#1S{\ensuremath{\Upsilon{\rm(#1S)}}}
\def\FourS {\Y4S}

\def\BzBzb {\ensuremath{B^0 \Bbar^0}} 
 
\def\Bbar {{\kern 0.18em\overline{\kern -0.18em B}}{}}
\def\Kstbar {{\kern 0.18em\overline{\kern -0.18em K^*}}{}}
\def\Dbar {{\kern 0.18em\overline{\kern -0.18em D}}{}}
\def\BB    {\ensuremath{B\Bbar}} 

\def\mus {\ensuremath{\,\mu{\rm s}}}

\def\pipi  {\ensuremath{\pi^+\pi^-}}

\def\bpsikl{\ensuremath{B^0 \to \jpsi \KL}}
\def\jpsi  {\ensuremath{{J\mskip -3mu/\mskip -2mu\psi\mskip 2mu}}}
\def\KL    {\ensuremath{K^0_{\scriptscriptstyle L}}} 
\def\degrees {\ensuremath{^{\circ}}}

\def\mum {\ensuremath{\,\mu\rm m}}

\def\Abar  {{\kern 0.18em\overline{\kern -0.18em A}}{}}

\def\CP{\ensuremath{C\!P}}

\def\Kbar  {{\kern 0.2em\overline{\kern -0.2em K}}{}}

\long\def\inst#1{\par\nobreak\kern 4pt\nobreak
    {\it #1}\par\vskip 10pt plus 3pt minus 3pt}

\begin{document}
{\pagestyle{empty}

\begin{flushright}
SLAC-PUB-\SLACPubNumber \\
\babar-PROC-\BABARPubYear/\BABARProcNumber \\
November, 2001 \\
\end{flushright}

\par\vskip 4cm

\begin{center}
\Large \bf \CP\ Violation in the $B^0$ meson system with BaBar
\end{center}
\bigskip

\begin{center}
\large 
J. Weatherall\\
Department of Physics and Astronomy \\
University of Manchester, Oxford Road \\
Manchester, M13 9PL, UK \\
(for the \lbabar\ Collaboration)
\end{center}
\bigskip \bigskip

\begin{center}
\large \bf Abstract
\end{center}
The \babar\ detector, at the PEP-II asymmetric $B$ Factory at SLAC collected
a sample of 32 million $B\Bbar$ pairs whilst operating at energies near the
$\Upsilon(4S)$ resonance between October 1999 and May 2001.  An study of 
time-dependent \CP-violating asymmetries in events where one neutral $B$
meson is fully reconstructed in a final state containing charmonium produced
the measurement $\sin 2\beta = 0.59 \pm 0.14 \, ({\rm stat}) \pm 0.05 
\, ({\rm syst})$, which constitutes an observation of \CP\ violation in the 
$B^0$ meson system at the 4$\sigma$ level.  Also presented are preliminary 
results from a study of \CP\ violation in the decays $B^0\to \pi^+\pi^-$ and
$B^0\to K^+\pi^-$.

\vfill
\begin{center}
Contributed to the Proceedings of the \\
Seventh Topical Seminar on the Legacy of LEP and SLC \\
8-11 October 2001, Siena, Italy
\end{center}

\vspace{1.0cm}
\begin{center}
{\em Stanford Linear Accelerator Center, Stanford University, 
Stanford, CA 94309} \\ \vspace{0.1cm}\hrule\vspace{0.1cm}
Work supported in part by Department of Energy contract DE-AC03-76SF00515.
\end{center}

\section{Introduction}

\CP\ violation was first observed in the decays of $K^0_L$ mesons in 
1964~\cite{Christenson:fg}.  It
was 37 years before this phenomenon was observed in another system, that of
the neutral $B$ mesons.  The effect arises from the existence of an
irremovable, \CP-violating phase in the three-generation CKM quark-mixing 
matrix~\cite{Cabibbo:yz}.
The $B^0$ system has an important advantage over the $K^0$ system in that the 
measurements of \CP-violating asymmetries provide a direct test of the 
Standard Model of electroweak interactions, free of corrections from strong
interactions, which arise when trying to interpret the results from $K^0_L$
decays.  The primary goal of \babar\ is to over-constrain the Unitarity 
Triangle through multiple, independent measurements of its sides and angles.
In this analysis, the angle $\beta$ is probed through a measurement of 
$\sin 2\beta$ using $b \to c \overline{c} s$ decays.

\section{PEP-II and \babar}

The PEP-II $B$ factory at SLAC comprises a pair of storage rings producing
asymmetric $e^+e^-$ collisions (9\,GeV $e^-$, 3.1\,GeV $e^+$) at a 
center-of-mass energy corresponding to the
mass of the $\Upsilon(4S)$ resonance (10.58\,GeV).  The $\Upsilon(4S)$ decays
almost exclusively to $B^+B^-$ or coherent $B^0\Bbar^0$ pairs.  The primary 
goal of the \babar\ detector for this analysis is to measure the time 
difference between the two meson decays, $\Delta t$.  
The asymmetric nature of the collisions gives a boost of $\beta\gamma=0.55$ 
which yields an average spatial vertex separation of $\approx250\mum$.

A detailed description of the \babar\ detector can be found 
in~\cite{Aubert:2001tu}.  The volume inside the 1.5T superconducting solenoid
consists of a five layer silicon vertex detector (SVT), a drift chamber
(DCH), a quartz Cerenkov detector (DIRC) and a CsI(Tl) crystal electromagnetic
calorimeter (EMC).  The instrumented flux return (IFR) outside the magnet 
comprises alternate layers of iron and resistive plate counters (RPCs).

\section{Exclusive $B$ reconstruction}

A sample of neutral $B$ mesons, $B_{\CP}$ have been fully reconstructed in 
their decays to final states of known \CP\ content: $J/\psi K^0_S$, $\psi(2S) 
K^0_S$, $J/\psi K^0_L$, $\chi_{c1} K^0_S$ and $J/\psi 
K^{*0}(K^{*0}\to K^0_S\pi^0)$.  
In addition there are also samples of $B$ decays to final states of 
definite flavour ($B_{\rm flav}$): 
$B^0\to D^{(*)-} \pi^+$, $D^{(*)-} \rho^+$, $D^{(*)-} 
a_1^+$ and $J/\psi K^{*0}(K^{*0}\to K^+\pi^-)$ as well as charged $B$ 
decays: $B^-\to D^{(*)0}\pi^-$, $J/\psi K^-$ and $\psi(2S) K^-$ 
(charge conjugate modes are implied throughout
this paper).  All selections have been optimized to give maximum sensitivity
to the final measurement.  Particle identification, mass (or mass difference)
and vertex constraints are used wherever applicable.  The resulting signal
yield for each mode is identified by using the kinematical variables 
$\Delta E = E^*_{B}-E^*_{beam}$ and $m_{ES} = \sqrt{E^{*2}_{beam}-p^{*2}_B}$
where $E^*_B$, $p^*_B$ are the center-of-mass energy and momentum of the 
reconstructed $B$ and $E^*_{beam}$ is the beam energy in the center-of-mass.  
In the case that an event has more than one $B$ candidate, only the one with 
the smallest $|\Delta E|$ is retained.  For each mode a signal region is 
defined as $\pm 3\sigma$ about (5.279,0) in the $m_{ES},\Delta E$ plane.  The 
$m_{ES}$ resolution is $\approx 3\,{\rm MeV}/c^2$, dominated by the spread of 
the beam energy.  The $\Delta E$ resolution is mode dependent and varies from 
about 10-33\,MeV.  Figure~\ref{fig:bcp} shows the $m_{ES}$ distributions for 
$B_{\CP}$ candidates containing a $K^0_S$ and the $\Delta E$ distribution for 
candidates containing a $K^0_L$.  The number of tagged events and the signal
purities, determined from fits to the $m_{ES}$ (all $K^0_S$ modes except 
$K^{*0}$) or $\Delta E$ ($K^0_L$ mode) distributions in data or from Monte
Carlo simulation ($K^{*0}$ mode) are shown in table~\ref{tab:sam}.

\begin{center}
\begin{figure}[htb]
\hspace*{1cm}
\includegraphics*[width=12cm]{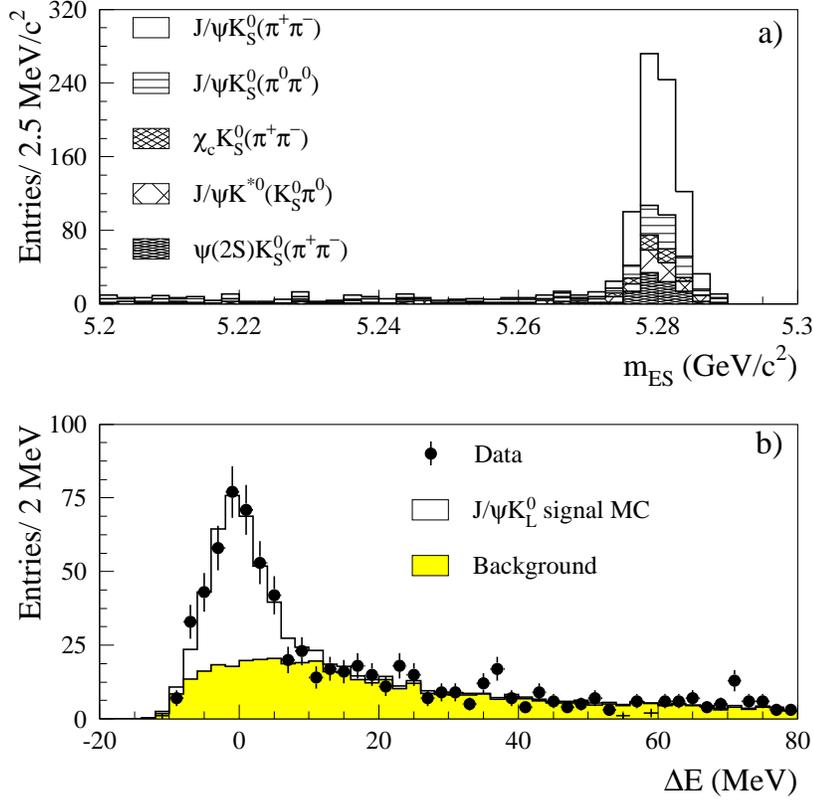}
\vspace*{-1.5cm}
\caption{a) Distribution of $m_{ES}$ for $B_{\CP}$ candidates having a $K^0_S$
in the final state; b) distribution of $\Delta E$ for $J/\psi K^0_L$ 
candidates.}
\label{fig:bcp}
\end{figure}
\end{center}

\begin{table*}[htb] 
\caption{ 
Number of tagged events, signal purity and result of fitting for \CP\ asymmetries in 
the full \CP\ sample and in 
various subsamples, as well as in the $B_{\rm flav}$ and charged $B$ control samples.  Errors are statistical only.}
\label{tab:sam} 
\vspace*{0.5cm}
\begin{center}
{\large
\begin{tabular}{cccc}
\hline \hline 
 Sample  & $N_{\rm tag}$ & Purity (\%) & $\sin 2\beta$ \\ \hline

$\jpsi\KS$,$\psitwos\KS$,$\chicone\KS$   & $480$        & $96$       &  0.56 $\pm$ 0.15   \\ 
$\jpsi \KL$ $(\eta_f=+1)$                & $273$        & $51$       &  0.70 $\pm$ 0.34   \\
$\jpsi\Kstarz ,\Kstarz \to \KS\piz       $& $50$         & $74$       &  0.82 $\pm$ 1.00  \\ 
\hline
 Full \CP\ sample                        & $803$        & $80$       &  0.59 $\pm$ 0.14   \\ 
\hline
\hline
\multicolumn{4}{l}{$\jpsi\KS$, $\psitwos\KS$, $\chicone\KS$  only  $(\eta_f=-1)$ }  \\
\hline
$\ \jpsi \KS$ ($\KS \to \pi^+ \pi^-$)    & $316$        & $98$       &  0.45 $\pm$ 0.18  \\ 
$\ \jpsi \KS$ ($\KS \to \pi^0 \pi^0$)    & $64$         & $94$       &  0.70 $\pm$ 0.50  \\ 
$\ \psi(2S) \KS$ ($\KS \to \pi^+ \pi^-$) & $67$         & $98$       &  0.47 $\pm$ 0.42   \\
$\ \chicone \KS $ ($\KS \to \pi^+ \pi^-$) & $33$         & $97$       &  2.59 $\pm$ $^{0.55}_{0.67}$ \\
\hline 
$\ $ {\tt Lepton} tags                   & $74$         &  $100$     &  0.54 $\pm$ 0.29   \\ 
$\ $ {\tt Kaon} tags                     & $271$        &  $98$      &  0.59 $\pm$ 0.20    \\ 
$\ $ {\tt NT1} tags                      & $46$         &  $97$      &  0.67 $\pm$ 0.45    \\ 
$\ $ {\tt NT2} tags                      & $89$         &  $95$      &  0.10 $\pm$ 0.74   \\ 
\hline 
$\ $ \Bz\ tags                           & $234$        &  $98$      &  0.50 $\pm$ 0.22     \\ 
$\ $ \Bzb\ tags                          & $246$        &  $97$      &  0.61 $\pm$ 0.22     \\ 
\hline\hline
$B_{\rm flav}$ non-\CP\ sample            & $7591$       & $86$       &  0.02 $\pm$ 0.04     \\
\hline 
Charged $B$ non-\CP\ sample        & $6814$       & $86$       &  0.03 $\pm$ 0.04     \\ \hline \hline
\end{tabular} 
}
\end{center}
\end{table*}

\section{Flavour Tagging}

Flavour tagging information is extracted from the other (partially 
reconstructed) $B$ in the event, $B_{\rm tag}$.  The coherent production of the
$B^0\Bbar^0$ pair ensures that the flavour of $B_{\CP}$ is exactly opposite
to that of $B_{\rm tag}$ at the time when $B_{\rm tag}$ decays, $t_{tag}$.  Each event
is assigned to one of four hierarchical, mutually exclusive tagging categories
or excluded from further analysis.  The {\tt Lepton} and {\tt Kaon} categories
contain events with high momentum leptons from semileptonic $B$ decays or with
kaons whose charge is correlated with the flavour of the decaying $b$ quark
(e.g. a positive lepton or kaon yields a $B^0$ tag).  The {\tt NT1} and 
{\tt NT2} categories are based on a neural network algorithm whose tagging 
power arises primarily from soft pions from $D^{*+}$ decays and from 
recovering unidentified isolated primary leptons.  The $B_{\rm flav}$ sample is
used to measure the tagging performance along with the $B_{\CP}$ events.  The 
figure of merit used is $Q_i = \epsilon_i(1-2w_i)^2$ where $\epsilon_i$ and 
$w_i$ are the efficiency and mistag fraction for category $i$.  The 
statistical error on $\sin 2\beta$ is proportional to $1/\sqrt{Q}$, where 
$Q = \sum Q_i$.  The efficiencies and mistag fractions for the four tagging 
categories are shown in table~\ref{tab:tag}.

\begin{table}[htb]
\caption{Efficiencies $\epsilon_i$ and average mistag fractions $w_i$  
extracted for each tagging category $i$ from a maximum likelihood fit to the 
time distribution for the fully reconstructed $B$ sample ($B_{\CP}+B_{\rm flav}$). 
Uncertainties are statistical only.}
\label{tab:tag}
\begin{center}
{\large
\begin{tabular}{lccc}
\hline \hline
Category & $\epsilon (\%)$ & $w (\%)$ & $Q (\%)$ \\ \hline
{\tt Lepton} & $10.9\pm0.3$ & $8.9\pm1.3$ & $7.4\pm0.5$ \\ 
{\tt Kaon} & $35.8\pm0.5$ & $17.6\pm1.0$ & $15.0\pm0.9$ \\ 
{\tt NT1} & $7.8\pm0.3$ & $22.0\pm2.1$ & $2.5\pm0.4$ \\
{\tt NT2} & $13.8\pm0.3$ & $35.1\pm1.9$ & $1.2\pm0.3$ \\ \hline
All & $68.4\pm0.7$ & & $26.1\pm1.2$ \\
\hline \hline
\end{tabular}
}
\end{center}
\end{table}  

\section{$\Delta t$ measurement and resolution}

The time difference between the two $B$ decays, $\Delta t$ = $t_{CP}-t_{tag}$
is determined by
first measuring their spatial separation $\Delta z = z_{\CP} - z_{tag}$.  This
is corrected on an event-by-event basis for the direction of the $B$ with 
respect to the $z$ direction in the $\Upsilon(4S)$ frame.  $z_{\CP}$ is 
determined from the charged tracks that constitute the $B_{\CP}$ candidate.
All other tracks in the event are fitted to a common vertex in order to 
calculate the $B_{\rm tag}$ decay position, $z_{tag}$.  Tracks from photon
conversions are removed.  Pairs of tracks compatible with the decay of a long
lived $K_S^0$ or $\lambda$ are replaced by their parent neutral pseudotrack.
The bias in the forward $z$ direction due to charm decays is reduced by 
removing the track with the largest contribution to the vertex $\chi^2$, if 
above 6 and iterating the fit until either no track fulfills this condition or
fewer than two tracks remain.  Knowledge of the beam spot location and beam 
direction is incorporated through the addition of a pseudotrack to the tagging
vertex, computed from the $B_{\CP}(B_{\rm flav})$ vertex and three-momentum, the
beam spot (with a vertical size of 10\,\mum) and the $\Upsilon(4S)$ momentum.
The total $\Delta z$ reconstruction efficiency is 97\%.  For 99\% of the 
reconstructed vertices the r.m.s. $\Delta z$ resolution is 180\,\mum, dominated
by the $B_{\rm tag}$ vertex.  An accepted candidate must have a converged fit for
the $B_{\CP}$ and $B_{\rm tag}$ vertices, an error of less than 400\mum\ on 
$\Delta z$ and a measured $|\Delta t| < 20 \, {\rm ps}$.

The $\Delta t$ resolution function for signal events is
represented as a sum of three Gaussian distributions.  All offsets are 
modelled to be proportional to the event-by-event error, $\sigma_{\Delta t}$,
which is correlated with the weight that the daughters of long-lived charm
particles have in the tag vertex reconstruction.  The `core' and `tail'
Gaussians have widths scaled by the event-by-event measurement error derived
from the vertex fits. 
A separate offset for the core distribution is allowed for each tagging 
category to account for small shifts caused by inclusion of residual 
charm decay products in the tag vertex.  The third Gaussian has a fixed width 
of 8\,ps and accounts for fewer than 1\% of events with incorrectly 
reconstructed vertices.  Identical resolution function
parameters are used for all modes, since the $\Delta t$ resolution is 
dominated by the $B_{\rm tag}$ vertex precision.  
Separate resolution function parameters
have been used for data collected in 1999-2000 and 2001, due to the 
significant improvement in the SVT alignment.

\section{Measuring $\sin2 \beta$}

\subsection{Improvements to the analysis}

There are several significant changes in this analysis~\cite{Aubert:2001nu} 
relative to the first
\babar\ $\sin 2\beta$ publication~\cite{Aubert:2001sp}.  The $B_{\CP}$ modes
$\chi_{c1} K^0_S$ and $J/\psi K^{*0}(K^{*0}\to K^0_S\pi^0)$ have been added.
Improvements in track and $K^0_S$ reconstruction efficiency in 2001 data 
produce a $\approx 30\%$ increase in the yields per luminosity unit.  Better
alignment of the tracking systems in 2001 data and improvements in the tag
vertex reconstruction algorithm have increased the sensitivity of the
measurement by an additional 10\%.  The purity of the sample has been 
increased by a reoptimization of the $J/\psi K^0_L$ selection.  In total, the
statistical power of the analysis is almost doubled with respect to that of
Ref.~\cite{Aubert:2001sp}.  The final $B_{\CP}$ sample contains about 640 
events, the $B_{\rm flav}$ sample has 7591 fully reconstructed $B^0$ events and 
the charged $B$ sample has 6814 fully reconstructed $B^{\pm}$ events.  

\subsection{The $\sin 2\beta$ fit}

The fit for $\sin 2\beta$ is based on the following framework.
Each event in the $B_{\CP}$ sample is examined for evidence that the $B_{\rm tag}$
decayed as a $B^0$ or $\Bbar^0$.  The decay distributions as a function of
time for events with either type of tag can be expressed in terms of a complex
parameter $\lambda$ that depends both on $B^0\Bbar^0$ mixing and the 
amplitudes describing $B^0$ and $\Bbar^0$ decay to a common final state $f$~\cite{Wolfenstein:sc}.  The distribution f$_+$(f$_-$) of the decay rate when the 
tag is a $B^0(\Bbar^0)$ is
\newpage
\begin{eqnarray}
{\rm f}_\pm(\, \deltat) = 
{\frac{{\rm e}^{{- \left| \deltat \right|}/\tau_{\Bz} }}{2\tau_{\Bz}
(1+|\lambda|^2) }}  \times  \left[ \ {\frac{1 + |\lambda|^2}{2}} \hbox to 2cm{}
\right. \nonumber \\ \left. \pm {\ \mathop{\cal I\mkern -2.0mu\mit m}}
\lambda  \sin{( \Delta m_{B^0}  \deltat )} 
\mp { \frac{1  - |\lambda|^2 } {2} } \cos{( \Delta m_{B^0}  \deltat) }   
\right], \nonumber 
\end{eqnarray}
where $\tau_{B^0}$ is the $B^0$ lifetime and $\Delta M_{B^0}$ is the 
oscillation frequency of $B^0\Bbar^0$ mixing.  The first oscillatory term is
due to interference between direct decay and decay after mixing.  A difference
between the $B^0$ and $\Bbar^0$ $\Delta t$ distributions or a $\Delta t$
asymmetry for either flavour tag is evidence for \CP\ violation.

In the Standard Model, $\lambda = \eta_f e^{-2i\beta}$ for 
$b\to c\overline{c}s$ decays containing charmonium where $\eta_f$ is the \CP\ 
eigenvalue of the state $f$ and $\beta = 
{\rm arg}[-V_{cd}V^*_{cb}/-V_{td}V^*_{tb}]$ is an angle of the Unitarity 
Triangle.  Thus, the time-dependent \CP-violating asymmetry is
\begin{eqnarray}
A_{\CP}(\deltat) &\equiv&  \frac{ {\rm f}_+(\deltat)  -  {\rm f}_-(\deltat) }
{ {\rm f}_+(\deltat) + {\rm f}_-(\deltat) } \nonumber \\%
&=& -\eta_f \stwob \sin{ (\Delta m_{B^0} \, \deltat )} , \nonumber
\label{eq:asymmetry}
\end{eqnarray}
where $\eta_f=-1$ for $J/\psi K^0_S$, $\psi(2S) K^0_S$ and $\chi_{c1} K^0_S$
and $+1$ for $J/\psi K^0_L$.  Due to the presence of even (L=0,2) and odd (L=1)
orbital angular momenta in the $J/\psi K^{*0}(K^{*0}\to K^0_S\pi^0)$ system
there can be \CP-even and \CP-odd contributions to the decay rate.  When the
angular information in the decay is ignored, the measured \CP\ asymmetry in
$J/\psi K^{*0}$ is reduced by a dilution factor $D_{\perp} = 1-2R_{\perp}$,
where $R_{\perp}$ is the fraction of the L=1 component.  It has been measured
as $R_{\perp} = (16 \pm 3.5)\%$ which, after acceptance corrections leads to an
effective $\eta_f = 0.65 \pm 0.07$ for this mode.

The $B_{\CP}$ and $B_{\rm flav}$ samples are used together in the unbinned
maximum likelihood fit for the extraction of $\sin 2\beta$.
A total of 45 parameters are varied in the fit, including 
$\sin 2\beta$ (1), the average mistag fraction $w$ and
the difference $\Delta w$ between $B^0$ and $\Bbar^0$ mistags for each tagging
category (8), parameters for the signal $\Delta t$ resolution (16) and 
parameters for background time dependence (9), $\Delta t$ resolution (3) and
mistag fractions (8).  The determination of the mistag fractions and signal
$\Delta t$ resolution function is dominated by the large $B_{\rm flav}$ sample.
Background parameters are governed by events with $m_{ES}<5.27 {\rm GeV}/c^2$.
As a result, the largest correlation between $\sin 2\beta$ and any linear 
combination of the other free parameters is only 0.13.  The $B$ lifetime and
mixing frequency are fixed at $\tau_{B^0} = 1.548\,{\rm ps}$ and
$\Delta M_{B^0} = 0.472 \, \hbar \, {\rm ps}^{-1}$ 
respectively~\cite{Groom:in}.  The value of $\sin 2\beta$ and the asymmetry
in the $\Delta t$ distribution were hidden using a blind analysis method 
(as described in~\cite{Aubert:2001sp}) until the event selection was optimized
and all other aspects of the present analysis were complete.

\begin{figure}[htbp]
\hspace*{2.5cm}
\includegraphics*[width=9cm]{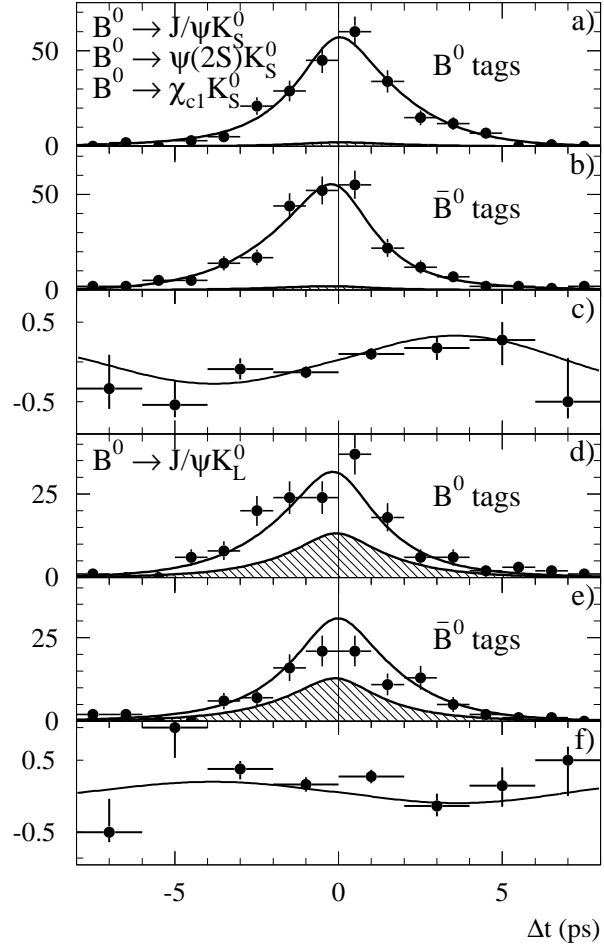}
\vspace*{-0.5cm}
\caption{a) Number of $\eta_f = -1$ candidates in the signal region a) with a 
$B^0$ tag, $N_{B^0}$ and b) with a $\Bbar^0$ tag, $N_{\Bbar^0}$ and c) the 
asymmetry $(N_{B^0}-N_{\Bbar^0})/(N_{B^0}+N_{\Bbar^0})$, as functions of
$\Delta t$.  The solid curves represent the result of the combined fit to all
selected \CP\ events and the shaded regions represent the background 
contributions.  Figures d)-f) contain the corresponding information for the 
$\eta_f = +1$ mode.  The likelihood is normalized to the total number of $B^0$
and $\Bbar^0$ tags.  The value of $\sin 2\beta$ is independent of the 
individual normalizations and therefore of the difference between the number
of $B^0$ and $\Bbar^0$ tags.}
\label{fig:res}
\end{figure}

Figure~\ref{fig:res} shows the $\Delta t$ distributions and $A_{\CP}$ as a 
function of $\Delta t$ overlaid with the likelihood fit results for the 
$\eta_f = -1$ candidates.  
The probability of obtaining a lower likelihood value than the one from the
fit, evaluated using a Monte Carlo technique, is 27\%.  The simultaneous 
fit to the full sample yields
\bd
\hspace{0.8cm}
\sin 2\beta = 0.59 \pm 0.14\,({\rm stat}) \pm 0.05 ({\rm syst}).
\ed

\subsection{Systematic errors and consistency checks}

The systematic error is dominated by the parametrization of the $\Delta t$
resolution function (0.03) - due in part to residual uncertainties in SVT 
alignment, possible differences in the mistag fractions between the $B_{\CP}$ 
and $B_{\rm flav}$ samples (0.03) and uncertainties in the level, composition and
\CP\ asymmetry of the background in the selected \CP\ events (0.02).

The large sample of reconstructed events allows a number of consistency checks,
including separation of the data by decay mode, tagging category and $B_{\rm tag}$
flavour.  The results of fits to these subsamples and to the samples of 
non-\CP\ decay modes (where no statistically significant asymmetry is found) 
are shown in table~\ref{tab:sam}.  



\section{Results from $B^0\to \pi^+\pi^-, K^+\pi^-$ decays}
A search for \CP-violating asymmetries in decays of $B^0$ to two light mesons
has been carried out on a sample of 33 million $B\Bbar$ 
pairs~\cite{Aubert:2001hk}.  Here one hopes to measure $\sin 2\alpha$ although 
the analysis is 
complicated by the possibility of penguin pollution leading to diagrams with
different weak and strong phases contributing to the same final state.  In 
this case,
one has to allow for direct \CP\ violation and in general $|\lambda|\neq 1$.
The fit is then complicated by the presence of a cosine term as well as the
coefficient of the sine term containing $\sin 2\alpha_{\rm eff}$, where 
$\alpha_{\rm eff}$ depends on the magnitudes and strong phases of the tree and 
penguin amplitudes.  The selected signal $B$ sample consists of 
$65^{+12}_{-11}$ $\pi\pi$, 217 $\pm$ 18 $K\pi$ and $4.3^{+6.3}_{-4.3}$ $KK$
events.  The results for \CP-violating asymmetries are summarized in 
table~\ref{tab:alp}.  Here $S_{\pi\pi}$ and $C_{\pi\pi}$ are respectively the 
coefficients of the sine and cosine terms in the expression for 
f$_{\pm}(\Delta t)$ and 
\bd
\hspace{1.5cm}
\mathcal{A}_{K\pi} \equiv \frac{N_{K^-\pi^+} - N_{K^+\pi^-}}
{N_{K^-\pi^+} + N_{K^+\pi^-}}.
\ed
Figure~\ref{fig:pipi} shows the $\Delta t$ distributions
and the asymmetry $\mathcal{A}_{\pi\pi}(\Delta t) = 
(N_{B^0}(\Delta t)-N_{\Bbar^0}(\Delta t))/
(N_{B^0}(\Delta t)+N_{\Bbar^0}(\Delta t))$ for tagged events which are 
enhanced in signal $\pi\pi$ decays.

\begin{center}
\begin{figure}[htb]
\hspace*{2.5cm}
\includegraphics*[width=10cm]{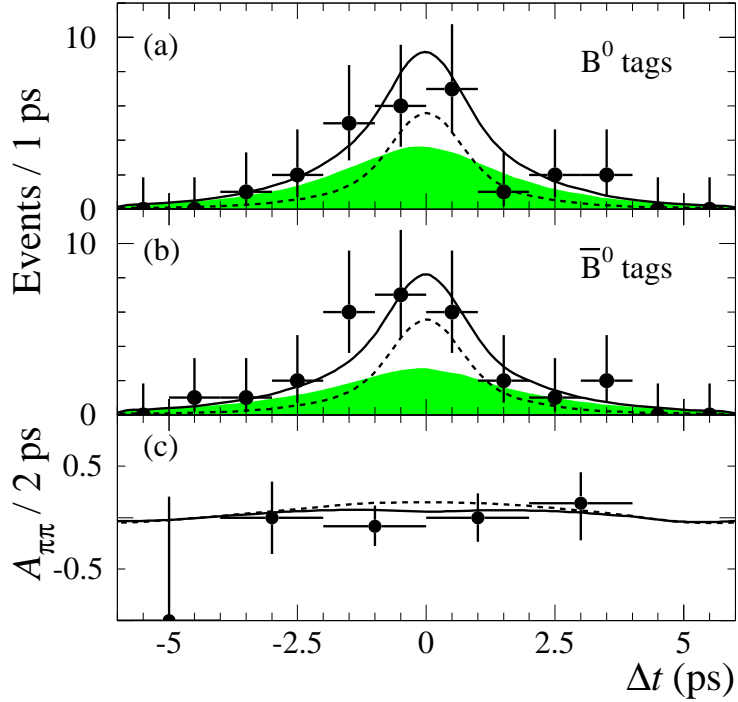}
\caption{Distributions of $\Delta t$ for events enhanced in signal $\pi\pi$
decays.  Figures (a) and (b) show events (points with errors) with 
$B_{\rm tag} = B^0$ or $\Bbar^0$.  Solid curves represent projections of the
maximum likelihood fit, dashed curves represent the sum of $q\overline{q}$ and
$K\pi$ background events and the shaded region represents the contribution from
signal $\pi\pi$ events.  Figure (c) shows $\mathcal{A}_{\pi\pi}(\Delta t)$ for
data (points with errors) as well as fit projections for signal and background
events (solid curve) and signal events only (dashed curve).}
\label{fig:pipi}
\end{figure}
\end{center}

\renewcommand{\arraystretch}{1.3}
\begin{table}[htb]
\caption{Central values and 90\% confidence level intervals for $S_{\pi\pi}$,
$C_{\pi\pi}$ and $\mathcal{A}_{K\pi}$.}
\label{tab:alp}
\begin{center}
{\large
\begin{tabular}{ccc}
\hline \hline
Parameter & Central Value & 90\% C.L. Interval \\ \hline
$S_{\pi\pi}$ & $0.03^{+0.53}_{-0.56} \pm 0.11$ & [-0.89,+0.85] \\
$C_{\pi\pi}$ & $-0.25^{+0.45}_{-0.47} \pm 0.14$ & [-1.0,+0.47] \\
$\mathcal{A}_{K\pi}$ & $-0.07 \pm 0.08 \pm 0.02$ & [-0.21,+0.07] \\
\hline \hline
\end{tabular}
}
\end{center}
\end{table} 
\renewcommand{\arraystretch}{1/1.3}

\section{Conclusions and outlook}
The measurement of $\sin 2\beta$ presented here establishes \CP\ violation in
the $B^0$ meson system at the 4.1$\sigma$ level, 37 years after its discovery
in the Kaon system.  The probability of obtaining this value of $\sin 2\beta$ 
or higher in the absence of \CP\ violation is $3\times 10^{-5}$.  This direct
measurement is consistent with the range implied by measurements and
theoretical estimates of the magnitudes of CKM matrix 
elements~\cite{CKMconstraints}.  By the 
summer of 2002, with a data sample of more than 100 million $B\Bbar$ pairs 
a measurement of $\sin 2\beta$ with a precision of less than 0.1 will be
possible.

In addition the search for \CP-violating asymmetries in the decays 
$B^0\to \pi^+\pi^-$ and $B^0\to K^+\pi^-$ looks promising and with
a similar amount of data should yield errors of $\approx0.3$ on 
the $B^0\to \pi^+\pi^-$ asymmetries.

\end{document}